\documentclass{iau}
\usepackage{graphicx}
\usepackage{hyperref}
\usepackage{multirow}

\title[Detecting complex sources in large surveys] 
{Detecting complex sources in large surveys using an apparent complexity measure}

\author[David Parkinson \& Gary Segal]   
{David Parkinson$^1$
 \and Gary Segal$^{2,3}$}

\affiliation{$^1$Centre for Theoretical Astrophysics, \\ Korea Astronomy and Space Science Institute,
776 Daedeok-daero,\\ Yuseong-gu, Daejeon 34055, Republic of Korea \\ email: {\tt davidparkinson@kasi.re.kr} \\[\affilskip]
$^{2}$School of Mathematics and Physics, University of Queensland,\\ St Lucia, Brisbane, QLD 4072, Australia\\
$^{3}$CSIRO Space and Astronomy, PO Box 76, Epping, 1710, NSW, Australia\\ email: {\tt gp.segal@gmail.com}}

\pubyear{2022}
\volume{368}  
\pagerange{119--126}
\jname{Machine Learning in Astronomy:
Possibilities and Pitfalls}
\editors{Ashish Mahabel, Christopher Fluke \& Jess McIver, eds.}
\begin{document}

\maketitle

\begin{abstract}
Large area astronomical surveys will almost certainly contain new objects of a type that have never been seen before. The detection of `unknown unknowns' by an algorithm is a difficult problem to solve, as unusual things are often easier for a human to spot than a machine. We use the concept of apparent complexity, previously applied to detect multi-component radio sources, to scan the radio continuum Evolutionary Map of the Universe (EMU) Pilot Survey data for complex and interesting objects in a fully automated and blind manner. Here we describe how the complexity is defined and measured, how we applied it to the Pilot Survey data, and how we calibrated the completeness and purity of these interesting objects using a crowd-sourced `zoo'. The results are also compared to unexpected and unusual sources already detected in the EMU Pilot Survey, including Odd Radio Circles, that were found by human inspection.
\end{abstract}

\firstsection 
\section{Introduction}

We are entering an unprecedented era of astronomical data. In the radio wavebands the pathfinders to the Square Kilometre Array will conduct wide and deep surveys. The Australian Square Kilometer Array Pathfinder (ASKAP) is predicted to detect $\sim70$ million radio galaxies with the Evolutionary map of the Universe (EMU) survey (\cite{EMUsurvey}), in comparison to the $\sim 2$ million extragalactic radio sources currently detected.
In the optical, the Legacy Survey of Space and Time (LSST), which will be conducted with the Vera Rubin Observatory (VRO) predicts a detection of 20 billion galaxies and a similar number of stars (\cite{LSST}), in comparison with the 300 million galaxies and 80 million stars in the Dark Energy Survey DR1 (\cite{DESDR1}), and the 1.6 billion objects in GAIA DR2 (\cite{GAIADR2}).

Many of these objects detected by these new surveys will be of a type never before seen by astrophysicists. This work focuses on a measure of morphological complexity that can be used to identify new and interesting objects. The approach does not requiring learning the specific features of past interesting observations, and as such, can be calibrated on a small sample and then applied to a much larger sample. This method differentiates itself from many existing outlier methods as it does not require the computationally intensive task of computing the distance between specific observations and the rest of the data in an abstract feature space. Rather complexity can be computed for a single observation and compared to a pre-determined threshold.

\section{Methodology}

\subsection{Coarse-grained complexity}

We introduce the coarse-grained complexity in \cite{2019PASP..131j8007S} based on the notions of \textit{effective complexity} (\cite{GellMann94, gell1996information}) and  \textit{apparent complexity} (\cite{Carroll}). The apparent complexity is a measure of the entropy $H$ of an object $x$ computed after applying a smoothing function $f$, expressed as $H (f (x))$. The Shannon entropy of a probability distribution P can be defined as the expected number of random bits that are required to produce a sample from that distribution:
\begin{equation}H(P) = -\sum_{x \in X} P(x)\log P(x) \,. \end{equation}

As discussed in \cite{Carroll, 2019PASP..131j8007S}, the Kolmogorov complexity $K (f (x))$  can be used as a proxy for the entropy of the smoothed function $H (f (x))$.  The Kolmogorov complexity of $x$ is the length of the shortest binary program $l(p)$, for the reference universal prefix Turing machine $U$, that outputs $x$; it is denoted as $K(x)$:
\begin{equation}
K(x) = {\rm min}_{p}\{l(p):U(p)=x\} \,. 
\end{equation}
While the Kolmogorov complexity is uncomputable, its upper bound can be reasonably approximated by the compressed file size $C (f (x))$ using a standard compression program such as \texttt{gzip}.

Intuitively a complexity measure should provide low values for random data that does not contain structure or regularities.  The coarse-grained complexity, like the apparent and effective complexity, can be defined as the compressed description length of regularities and structure after discarding all that is incidental.  The coarse-grained complexity achieves this by applying a smoothing function $f$ to the input $x$ (which removes fine-grained noise while preserving the coarse-grained structure of the image) and calibrating this so that complexity values correctly partition data that has been expertly labelled  (i.e. by human astronomers).
\subsection{Complexity Scan}

\begin{figure}[htp]
\begin{center}
 \includegraphics[width=3.0in]{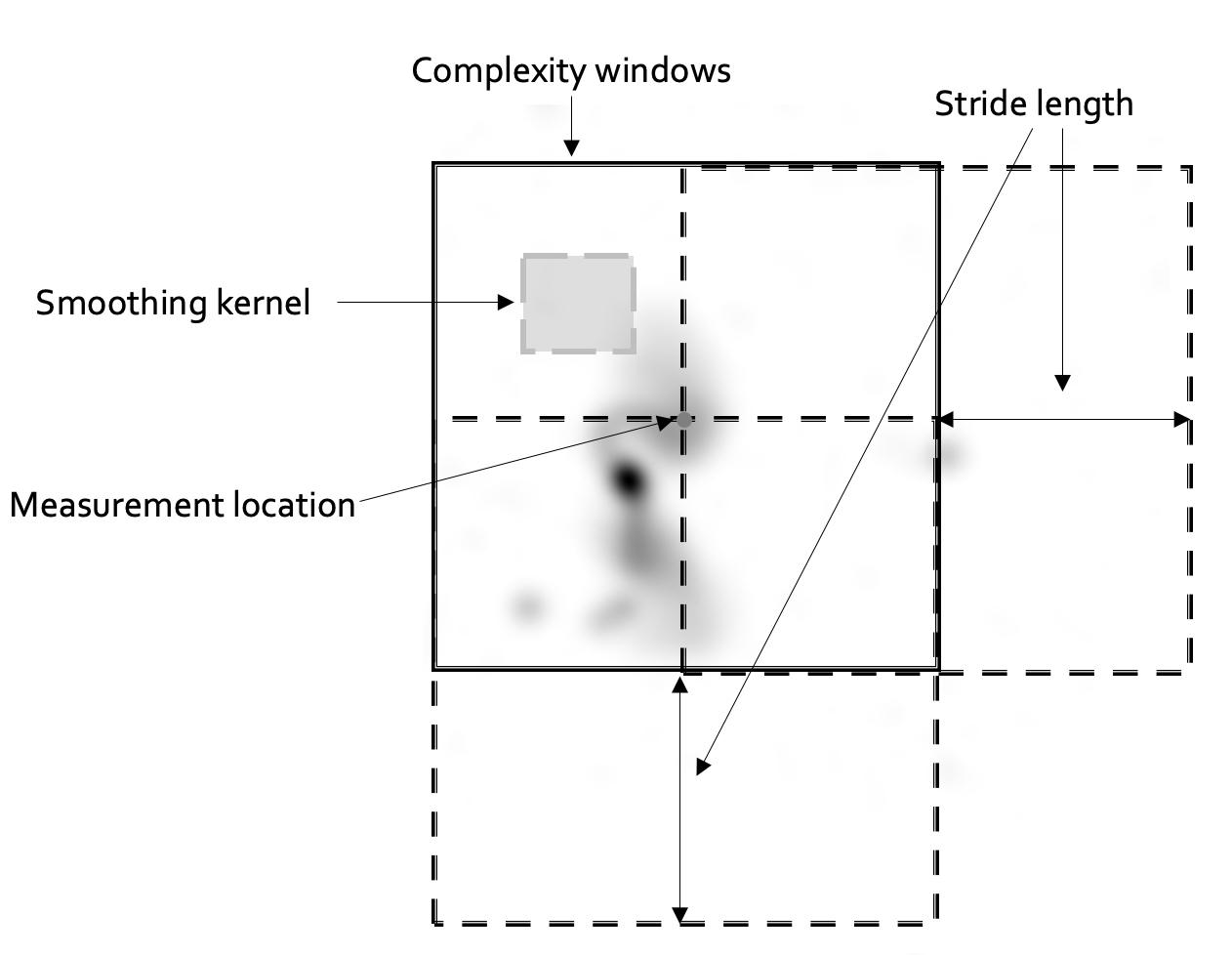} 
 \caption{\label{fig:stride} Illustration complexity scan parameters, including: the frame size (side length of solid black square), stride length (offset between position of solid black square and dashed black squares) and smoothing kernel size (grey shaded rectangle).} 
   \label{fig1}
\end{center}
\end{figure}

Source detection algorithms may make assumptions, and so miss interesting things. Instead, we scan the region, computing the course-grained complexity within a sliding frame (square aperture, as shown in Fig.~\ref{fig:stride}). 

An important feature of the scan method is that frames are sampled without making any assumptions as to whether the frame contains a conventionally detected source or not (that is without using a source extraction tool or existing catalogue data). This helps reduce the risk of producing a sample that is biased towards  preconceived notions of what is interesting, referred to as expectation bias by \cite{norris2017} and \cite{Robinson87}. This process builds up a ‘heat map’ of complexity values intended to assist with the identification of the unexpected in new and large data with the goal of new scientific discoveries and surprise.

\section{Data}

\subsection{EMU Pilot Survey}

The Pilot Survey of the Evolutionary Map of the Universe (EMU-PS) was observed at 944MHz using the Australian Square Kilometre Array Pathfinder (ASKAP) telescope. The ASKAP telescope consists of 36 12-metre antennas spread over a region 6-km in diameter at the Murchison Radio Astronomy Observatory in Western Australia. EMU-PS covers 270 square degrees of an area covered by the Dark Energy Survey at a spatial resolution of $\sim$ 11–13 arcsec (\cite{2021PASA...38...46N}).

While the primary goal of the EMU Pilot Survey is to test and refine observing parameters and the strategy for the main survey, the pilot in itself presents opportunity for new discoveries. Experience has shown (\cite{norris2017}) that whenever we observe the sky to a significantly greater sensitivity we make new discoveries. This goal has already been demonstrated through the successful identification of a new class of radio object, odd radio circles (ORCs, \cite{norris2021a,10.1093/mnrasl/slab041, norris22}). 

\subsection{Anomaly Zoo}

Truth labels are required to evaluate the effectiveness of alternative complexity thresholds for partitioning anomalous sources. To provide truth labels for the frames produced by the complexity scan we ran a project on the \url{Zooniverse.org} platform, titled \textit{``Anomaly in the EMU Zoo''} (hereafter zoo), requesting expert astronomers to evaluate an unbiased sample of frames sub-sampled from the EMU-PS scan. Consensus from the zoo labels was then used to evaluate the Recall and Precision associated with prospective partition boundaries. 

Expert volunteers were approached from within the Evolutionary Map of the Universe Survey Project and at the SPARCS 2021 conference. A sub-sample of 1627 frames from the EMU-PS scan ($n_{\mathrm{total}}$=365,000) were presented to volunteers for classification through the Zooniverse project. 44 volunteers participated in the project, with 10 of these classifying more than 500 frames each. 

The zoo asked the expert volunteers to evaluate frames sub-sampled from the EMU-PS Scan and to select an option that best describes the most interesting radio sources in each frame before moving on to the next. The four options presented for selection were:
\begin{itemize} 
\item No sources/just noise
\item One or more simple sources/unrelated simple sources
\item At least one complex source/sources with multiple components
\item Contains something unexpected/Anomaly
\end{itemize}

The zoo distinguished between complex and extended sources with multiple components, and sources that were deemed by the volunteers to be truly unexpected or anomalous. Only frames converging on a label through majority consensus were used to evaluate the effectiveness of the complexity to identify anomalous sources. Those frames from the zoo sample where majority consensus was not reached were excluded from the evaluation.

\section{Results}

\subsection{Complexity Heat Map}

\begin{figure*}
\begin{center}
\includegraphics[width=8cm]{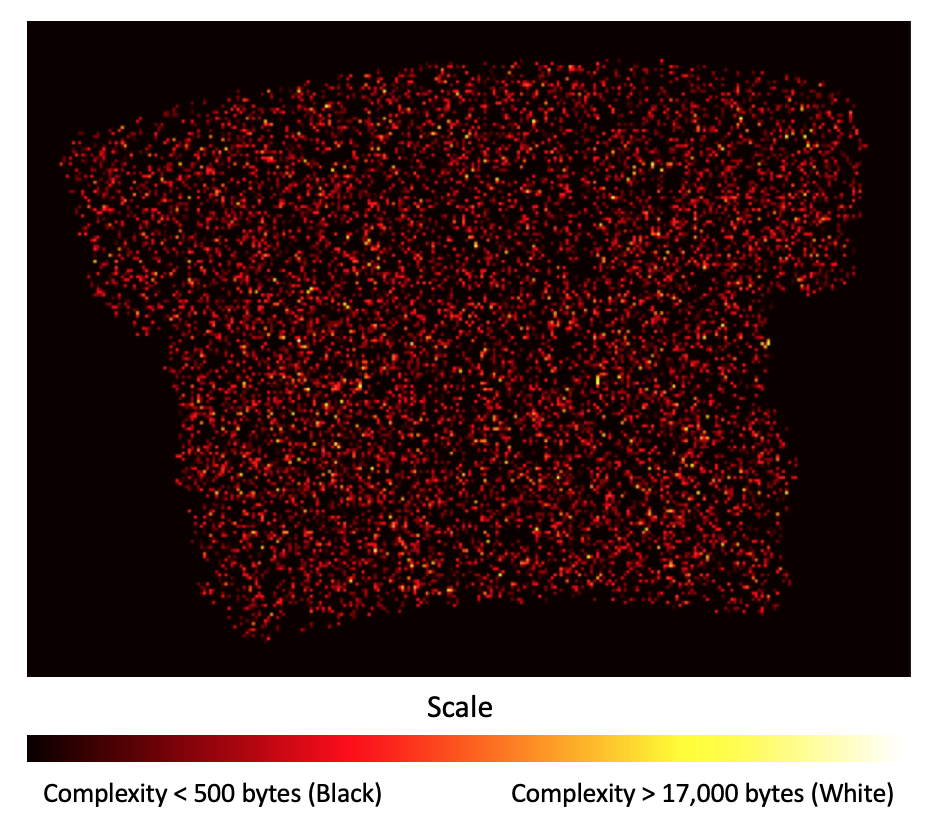}
\caption{\label{fig:hmap} A complexity `heat map' of the EMU-PS region, where the complexity of the frame is indicated by the brightness of the pixel. While many of the frames are red (grey) coloured, indicating the presence of a simple source, there are a few yellow (light grey) or white pixels, indicating more complex sources or low-surface brightness structures.} 
\end{center}
\end{figure*}

We performed a complexity scan of the  EMU-PS data using the method described. Complexity values are shown in Fig.~\ref{fig:hmap} as a heat map overlaid on the EMU-PS field.

\begin{figure}
\includegraphics[width=8cm]{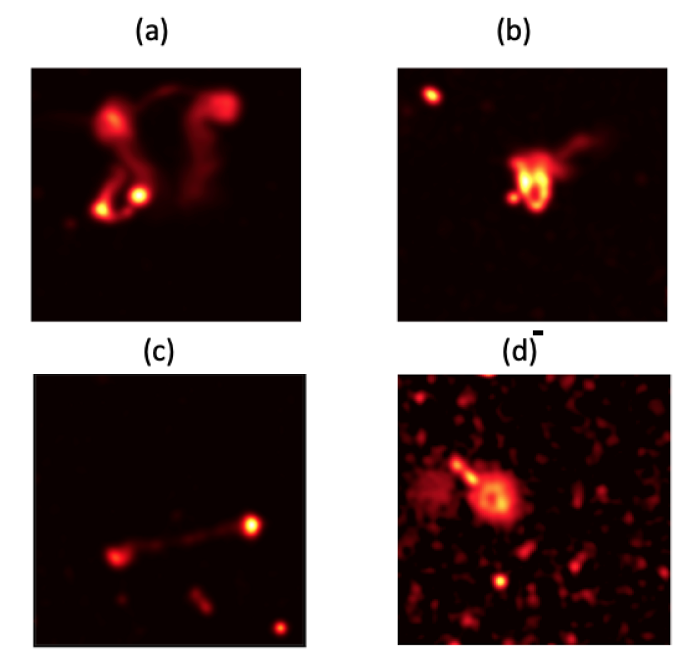}
\caption{\label{fig:examples} Examples of objects found within frames sampled above the 99th percentile complexity value from the EMU-PS Complexity Scan. These examples illustrate the breadth of objects found within frames in the complexity tail including, (a) the unusual source PKS 2130-538 (\cite{Otrupcek}) known as the Dancing Ghosts (\cite{2021PASA...38...46N})
, (b) a bright wide angle tail source on 2MASX J21291901-5053040 in cluster Abell 3771, (c) a previously unreported Giant Radio Galaxy  with projected linear length of 1.94 Mpc (DES J221640.02-625241.6), (d) two odd radio circles, EMU PD J205842.8–573658 (ORC2) and EMU PD J205856.0-573655 (ORC3).
}
\end{figure}

Visual inspection of a sub-sample of frames shows that the high-complexity value tail of the distribution comprises unusual, complex and extended objects with examples shown in Fig.~\ref{fig:examples}. These include peculiar Tailed Radio Galaxies, Fanaroff-Riley Class I (FRI) and Class II (FRII) type sources with more inflated lobes than that of average such objects, Odd Radio Circles, and Giant Radio Galaxies including a previously unreported Giant Radio Galaxy with projected linear length of 1.94 Mpc.

\subsection{Anomaly Zoo Outcomes}

We use truth labels determined from the zoo to evaluate the effectiveness of alternative partitions for identifying anomalous objects. As an immediate benefit, an effective partition can be used to identify frames containing complex structures and unusual objects from the EMU-PS data and build an anomaly catalogue. An effective boundary can also be used when analysing new, and even larger surveys, including the subsequent full EMU survey which is anticipated to capture approximately 40 million sources.

Through sub-sampling of the EMU-PS Scan, the sample size for the zoo was limited to 1,627 frames. An enrichment sample was included to provide better representation of the type of observations found within frames beyond the 99.5th percentile. Classification counts for each of the zoo classes are shown in Tab.~\ref{tab:zooclassification}. This table includes the anomaly counts both before and after the inclusion of the enrichment sample.

\begin{table}
\begin{tabular}{l c c c}
\hline
Classification & Count:  & Count: & Count: \\
 & Non-bias sample & Enrichment & Combined Sample \\
 & (n=1528) & (n=99) & (n=1627)  \\
\hline 
Anomalous& 7 & 17 & 24\\
Complex& 366 & 57 &423 \\ 
Simple& 1059 & 12 &1071 \\ 
No source& 31 & 0 &31 \\ 
No consensus& 65 & 13 &78 \\ 
\hline
\end{tabular}
\caption{\label{tab:zooclassification} Classification counts by consensus for each of the options provided in the EMU-PS Anomaly in the EMU Zoo. The enrichment sample supplements the zoo sub-sample with frames measuring complexity above the 99.5th percentile.} 
\end{table}

Only frames converging on a label through majority consensus were used to evaluate the effectiveness of the complexity to identify anomalous sources. Those frames from the zoo sample where majority consensus was not reached were excluded from the evaluation. 99.8\% of the zoo sample frames received evaluations from 3 or more volunteers and 94\% from 5 or more. We used the criterion  that only frames receiving 5 or more evaluations were used to evaluate consensus, and be given a reliable `truth' label. These restrictions were imposed to avoid the results being impacted by outlier evaluations that differed from the majority of expert volunteers.

\begin{figure*}
\centering
\includegraphics[width=3.5in]{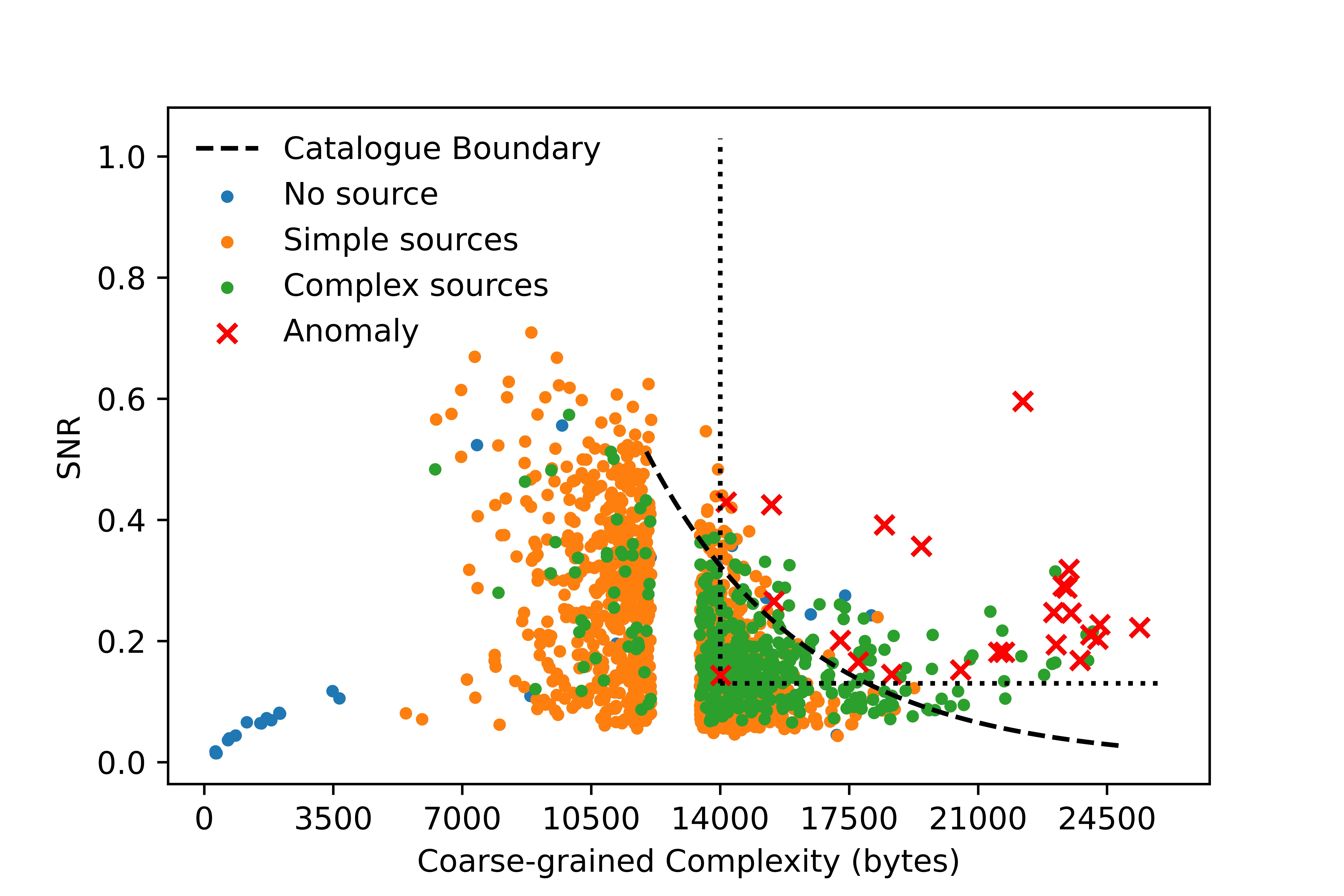}
\caption{\label{fig:Enriched_Scatter} Scatter plot showing the coarse-grained complexity and the signal-to-noise ratio (SNR), and the relationship between them, along with the classification of frames derived through consensus from the zoo (Anomaly in the EMU Zoo).}
\end{figure*}

We partition the data based on complexity and signal-to-noise ratio (SNR) values and use this to define a function based catalogue boundary (shown in Fig.~\ref{fig:Enriched_Scatter}). 

\section{Summary}

The coarse-grained complexity can be used as a tool for identifying unusual and complex objects.  We apply the method to new  EMU-PS data (365,000 sampled frames containing at least the ~220,000 Selavy catalogue sources) to identify and segment unusual sources (i.e. anomalies). 

We used a Zooinverse project to produce crowd-sourced labels to evaluate the effectiveness of the approach and we identified an effective anomaly partition using the coarse-grained complexity and SNR values. 

Results demonstrate the ability of the coarse-grained complexity to single out regions of the sky that contain complex and unusual sources, in a manner that can be computed at worst-case linear time complexity without reference to existing catalogue data. We propose the complexity measure as useful tool for identifying regions of interest in subsequent large and deep radio continuum surveys.

\end{document}